\documentstyle[11pt,aasms]{article}
 
\def\lta{{\>\rlap{\raise2pt\hbox{$<$}}\lower3pt\hbox{$\sim$}\>}}
\def\gta{{\>\rlap{\raise2pt\hbox{$>$}}\lower3pt\hbox{$\sim$}\>}}
 
\begin{document}

\title{Spiral Galaxies with HST/NICMOS. II. Isophotal Fits and Nuclear
Cusp Slopes\footnote{Based on observations with the NASA/ESA Hubble
Space Telescope, obtained at the Space Telescope Science Institute,
which is operated by the Association of Universities for Research in
Astronomy, Inc. (AURA), under NASA contract NAS5-26555 }}

\author{Marc Seigar\footnote{Joint
Astronomy Centre, 660 N. A'ohoku Place, Hilo, HI 96720} \affil{Sterrenkundig Observatorium, Universiteit
Gent, Krijgslaan 281, B-9000 Gent, Belgium \\ Space Telescope Science
Institute, 3700 San Martin Drive, Baltimore MD 21218}}

 \author{C.\ Marcella Carollo\footnote{Visiting Astronomer, The Johns
Hopkins University, Department of Physics and Astronomy, Baltimore MD
21210}} \affil{Columbia University, Department of Astronomy, 538 W.\
120$^{th}$ St., New York, NY 10027}

\author{Massimo Stiavelli} \affil{Space Telescope Science Institute,
3700 San Martin Drive, Baltimore MD 21218}

\author{P.\ Tim de Zeeuw} \affil{Sterrewacht Leiden, Postbus 9513, 2300
RA Leiden, The Netherlands}
 
\author{Herwig Dejonghe} \affil{Sterrenkundig Observatorium,
Universiteit Gent, Krijgslaan 281, B-9000 Gent, Belgium}

\begin{abstract}

We present surface brightness profiles for 56 of the 78 spiral
galaxies observed in the HST/{\tt NICMOS/Camera-2} F160W snapshot survey
introduced in Carollo et al. (2001; paper I), as well as surface
brightness profiles for 23 objects out of the 41 that were also
observed in the F110W filter. We fit these surface brightness profiles
with the Nuker law of Lauer et al.\ (1995), and use the smooth
analytical descriptions of the data to compute the average nuclear
stellar cusp slopes $\langle \gamma \rangle$ in the 0.1$''$-0.5$''$
radial range.  Our main result is the startling similarity between the
nuclear stellar cusp slopes $\langle \gamma \rangle$ in the
near-infared compared to those derived in the visual passband.  This
similarity has several implications: (1.)  Despite the significant
{\it local} color variations that are found in the nuclear regions of
spirals and that are documented in paper I, there are typically little
or no optical-NIR {\it global} color gradients, and thus no global
stellar population variations, inside $\sim 50$-100 pc from the
nucleus in nearby spirals.  (2.)  The large observed range of the
strength of the nuclear stellar cusps seen in the HST optical study of
spiral galaxies reflects a physical difference between galaxies, and
is not an artifact caused by  nuclear dust and/or recent star
formation. (3.) The dichotomy between $R^{1/4}$ bulges, with
steep nuclear stellar cusps $\langle \gamma \rangle \sim 1$, and
exponential bulges, with shallow nuclear stellar cusps $\langle
\gamma \rangle < 0.3$, is also not an artifact of the effects of dust
or recent star formation. (4) The presence of a surrounding massive
disk appears to have no effect on the rise of the stellar density
distribution within the innermost hundred pc of the $R^{1/4}$
spheroids.\\ These results imply a breakdown within the family of
exponential bulges of the nuclear versus global relationships that
have been found for the $R^{1/4}$ spheroids. Such a breakdown is
likely to have significant implications concerning the formation of
exponential bulges and their connection with the $R^{1/4}$ spheroids.

\end{abstract}

{\bf keywords:}  Galaxies: Spirals --- Galaxies: Structure ---
Galaxies: Nuclei --- Galaxies: Bulges

\section{Introduction}

\noindent
This is the second of two papers in which we report the results of a
{\it Hubble Space Telescope} (HST) {\tt NICMOS} snapshot survey of the nuclear
regions of nearby spirals in the near-infrared (NIR).  As discussed in
more detail in the companion paper (paper I; Carollo et al.\ 2001a),
we have previously conducted an optical HST snapshot survey of the same targets
with {\tt WFPC2}. The optical survey has shown that the innermost few
hundred parsecs of spirals contain nuclear bars, spiral-, ring-like or
irregular distributions of recent star formation and dust, and, in
many cases, central spatially-resolved, photometrically-distinct
`nuclei' significantly brighter than their surroundings (Carollo et
al.\ 1997a; Carollo et al.\ 1998, C98; Carollo \& Stiavelli 1998,
CS98; Carollo 1999).  The optical survey has furthermore shown that in
the $V$-band massive early-type spiral bulges have steep nuclear
stellar cusps, i.e., cusp slopes $\gamma \gta 1$ in the power-law
approximation of the surface brightness profile $I(r) \rightarrow
r^{-\gamma}$ as $r \rightarrow 0.1''$. This is similar to what is
observed for elliptical galaxies of comparable luminosities (Jaffe et
al.\ 1994; Lauer et al.\ 1995; Forbes, Franx, \& Illingworth 1995;
Carollo et al.\ 1997b,c; Faber et al.\ 1997). In contrast, the
later-type systems often host bulgelike central structures with an
exponential rather than $R^{1/4}$-law radial profile (`exponential
bulges', see also Courteau et al.\ 1996 and references therein), and
show shallow nuclear stellar cusps $\gamma \lta 0.3$ and nuclear
stellar densities at least a factor 10 lower than those that have been
inferred for the massive $R^{1/4}$ bulges (CS98).

The visual-band {\tt WFPC2} results are potentially severely affected
by dust and recent star formation in the central regions of spirals;
the NIR window is less sensitive to such `polluting' factors, and thus
is much better suited than the visual passband for investigating the
underlying older stellar populations. We have therefore followed up
the {\tt WFPC2} survey with  a {\tt NICMOS/Camera-2} ($0.075''$/pixel
scale) snapshot survey in the $H$ (F160W) passband and, for a fraction of the
galaxies, also in the $J$ (F110W) band.  In paper I we described the
properties of the {\tt NICMOS} sample and the procedure adopted for
the basic reduction of the images, and discussed the NIR images and
the optical-NIR color maps of spirals at HST resolution. There we also
described the methodology used to measure the sizes and luminosities
of the photometrically-distinct nuclei, and discussed the statistical
NIR and optical-NIR properties of the nuclear regions of spirals,
including the distinct nuclei, as a function of Hubble type.  Jointly
with the {\tt WFPC2} data, the {\tt NICMOS} data also provide nuclear
optical-NIR colors, i.e., information on the nuclear stellar
populations. These are important for understanding the evolutionary
history of the central regions of spirals. We have presented in
Carollo et al.\ (2001b) a discussion of the optical-NIR HST colors of
bulges and nuclei in our sample.

In this paper we present the isophotal fits performed on the {\tt
NICMOS/Camera-2} images  in order to investigate
the nuclear stellar structure that underlies the distinct nuclei and
the nuclear dust patches and bar-, ring- and arm-like
features. Specifically, we {\it (i)} describe the methodology adopted
to perform the isophotal fits, and present the nuclear surface
brightness profiles (\S 2; the brightness profiles
are available in electronic format upon request); 
{\it (ii)} describe the results of
modeling the nuclear surface brightness profiles with the analytical
law introduced by Lauer et al.\ (1995), and use these analytical
descriptions to derive the strength of the nuclear stellar cusp slopes
in the NIR (\S 3); {\it (iii)} present and discuss the results of our
analysis as a function of global galactic properties (\S 4). We
summarize in \S 5. Consistently with the previous papers, here we
adopt $H_o=65$ km/s/Mpc.

\section{Derivation of the Surface Brightness Profiles}

\noindent
The derivation of the surface photometry was carried out in {\tt IRAF}
by using the `ellipse' task. For all the galaxies we first performed
the isophotal fits without masking the knots of star formation and the
dust lanes frequently present in the images.  For the galaxies with
strong star formation or dust patterns, these fits turned out to be a
very poor representation of the smooth distribution of the underlying
stellar light which we seek to parametrize.  For these objects we
therefore rederived the final isophotal fits after masking out stars
and other knot-like sources such as HII regions and strong dust
lanes. In some galaxies the distinct nucleus is slightly offset from
the center of the isophotes (paper I); in these cases the nucleus was
simply masked out before performing the isophotal fit.  The isophotal
analysis could be derived for 56 galaxies in the $H$-band and for 23
objects in the $J$-band; in the remaining galaxies, the images had too
little S/N, or the morphology was too irregular, for performing a
reliable isophotal fit.  We chose not to carry out any Point Spread
Function (PSF) deconvolution of the images before fitting the
isophotes, but to convolve instead the analytical models of the
surface brightness profiles described in \S 3 with the {\tt NICMOS}
PSF before comparing them to the observed data points.  Typically, no
suitable star was present in the {\tt NICMOS} Camera-2 field of view
of the target galaxies; appropriate PSFs for the F110W and F160W
filters were thus derived with the Tinytim software package (Krist
1997).  The approach of convolving the models with the PSF rather than
deconvolving the images helps to minimize the effects of a simulated
PSF, which does not perfectly represent the true PSF.

Absolute photometric calibration in the AB magnitude system was
obtained by applying zero-point corrections. These were taken to be
equal to $Z_{{\tt NIC2,F110W}}=23.25$ and $Z_{{\tt NIC2,F160W}}=23.11$
for the the {\tt F110W} and the {\tt F160W} filter, respectively,
so that:
\begin{equation}
m_{\lambda}=-2.5\log_{10}(\hbox{\rm counts/sec}) + 
             Z_{{\tt NIC2},\lambda}+5\log_{10}(0.075)
\end{equation}
where the $5\log_{10}(0.075)$ term accounts for the pixel size. The
calibrated surface brightness profiles were corrected for Galactic
extinction by using the values published by Burstein \& Heiles (1984).
Figure 1 a shows the derived 56 $H$-band surface brightness profiles,
and Figure 1 b shows the 23 similar measurements in the $J$-band.

\section{Analytical Fits to the Surface Brightness Profile}

\noindent
A few analytical expressions have been proposed to date to describe
the surface brightness profiles in the inner regions of galaxies.  For
spiral galaxies, the presence of photometrically-distinct nuclei leads
to the question of whether it is best to include these subcomponents
in the analytical description of the nuclear surface brightness
profiles.  CS98 tested whether the optical nuclear light profiles of
spirals could be fitted by a standard exponential profile with the
addition of a second component describing a steep nuclear cusp, i.e.,
by:
\begin{equation}
\label{analaw}
I(R)=I_{0}\left(1+\frac{R_{c}}{R}\right)^{\gamma'}\exp{(-R/R_{s})}.
\end{equation}
For radii $R$ much smaller than the ``cusp'' radius $R_c$, this
profile describes a cusp with slope $\gamma'$. $I_0$ is the central
brightness of the exponential component, and $R_s$ the exponential
scalelength. CS98 found no physically meaningful fit for any of the
galaxy light-profiles with the exception of one, this demonstrating that
'steep-cusp exponentials' could exist, and that a simple fitting
procedure of the analytical light profile [Eq.\ \ref{analaw}] would
identify them.  The lack of a good fit for most cases indicates
however that the distinct nuclei present in several galaxies cannot be
described as a steepening of the light profile: these sources are a
photometrically-distinct component on top of the underlying galaxy
light.  Thus, a central nucleus that sits right on top on the center
of the isophotes needs to be `excluded' from the parametrization.  For
the latter we adopted the expression introduced by Lauer et al.\
(1995) for the elliptical galaxies (`Nuker law') and used by CS98 for
the $V$-band light profiles of spirals. This choice allows us to
directly compare the optical and NIR results obtained for the spiral
galaxies, as well as systematic differences between spiral and
earlier-type galaxies.  The Nuker law reads:
\begin{equation}
\label{physlaw}
I(r)=2^{(\beta-\gamma)/\alpha}I_{b}
\left(\frac{R_{b}}{r}\right)^{\gamma}\left[1+
\left(\frac{r}{R_{b}}\right)^{\alpha}\right]^{(\gamma-\beta)/\alpha}.
\end{equation}
The parameter $\gamma$ measures the steepness of the rise of the
profile toward the very center, i.e., the value of $\gamma$ in
$I(r)\sim r^{-\gamma}$ as $r\rightarrow0$; $R_b$ is the break radius
where the profile flattens in some cases to a shallower slope;
$\beta$ is the slope of the outer profile; $\alpha$ controls the
sharpness of the transition between inner and outer profile; $I_b$
is the surface brightness at $R_b$.  

In Figures 1 a and b we show the PSF-convolved analytical best fits
superimposed on the $H$- and $J$-band surface brightness profiles,
respectively.  For each galaxy fitted in the $H$-band, Table 1 lists
the radial range of the reported best fit and the corresponding best
fit values for $R_b$, $\alpha$, $\beta$, $\gamma$ and $\mu_b=-2.5 \log
I_b$. Table 2 lists the similar parameters for the $J$-band fits.

Since the intrinsic form of the light profile of the nuclei is not
known, there is clearly an uncertainty associated with removing the
nucleus contribution from a galaxy light profile. We quantified this
uncertainty by performing several fits to the light profile of the
same galaxy after varying the inner radial cutoff. In Tables 1 and 2
the listed error bars for the parameters refer to the largest between
the following two quantities: (1) the formal errors of the reported
best fits, and (2) the differences between values obtained from the
various fits performed within the different radial ranges.

\subsection{The average logarithmic slope $\langle \gamma \rangle$}

\noindent
For the galaxies where a continuous rise of the light profile is
observed down to the resolution of the data, the fits with the
Nuker-law are indeterminate; nonetheless, they still yield a smooth
analytical representation of the data that is unaffected by PSF and
central distinct-nucleus effects. Thus, these analytical fits remain
best suited for deriving a global representation of the galactic
stellar light in the NIR. To this purpose we computed, from the smooth
curves provided by the best fits of equation (\ref{physlaw}) to the
$J$ and $H$ light profiles, respectively\, the average logarithmic
slopes $\langle \gamma^{J} \rangle$ and $\langle \gamma^{H}
\rangle$. The radial range adopted for deriving these average
logarithmic slopes was $0.1''-0.5''$, i.e., coincident with the range
adopted by CS98 for deriving an analogous quantity from the optical
light profiles of spiral galaxies. This radial range is equivalent to
a physical scale $\sim 50$-100 pc at the average galaxy distance of
$\approx 25$ Mpc.

Computing an average slope within the $0.1''$-$0.5''$ radial range has
the advantage of producing a description of the nuclear profile that
is independent of its particular parametrization. On the other hand,
the average nuclear slopes are more sensitive to distance effects than
is the asymptotic slope $\gamma$.  In galaxies with otherwise
identical surface brightness profiles, the more the surface brightness
profiles deviate from a pure power-law, the more the average slope
values vary with distance. However, this is not a problem in our data
set, since most of our galaxies typically lie within $\sim$20
$h_{65}^{-1}$ Mpc, and none lies beyond $\sim$40 $h^{-1}$ Mpc.

The values of $\langle \gamma^{H} \rangle$ and $\langle \gamma^{J}
\rangle$ are reported also in Tables 1 and 2. The associated error
bars refer to the largest between (1) the formal errors of the logarithmic
fits, and (2) the standard deviations obtained when estimating $\langle
\gamma \rangle$ by using all the fits derived using different radial
cut offs and that provided a physically meaningful description of the
given light profile. These error bars are typically smaller than $\sim
0.2$.

\section{Results \& Discussion}

\noindent
The almost ubiquitous presence of dust and often of star formation
knots in the central regions of spiral galaxies implies that a
significant fraction of the optical light from the nuclei of spirals
could either be absorbed by dust, similarly to what was suggested for
the ``shallow-core'' elliptical galaxies by Silva \& Wise (1996), or
be dominated by recently formed stars. This means that in principle
significant differences in nuclear stellar cusp slopes could be
expected when deriving these quantities by using optical and
near-infrared data. The latter are much less affected by recent star
formation, dust absorption and reddening in the nucleus, and thus
provide a better representation of the rate of density increase in the
underlying stellar populations with respect to the bluer wavelengths).
In Figure 2 a we plot the comparison between the average nuclear cusp
slopes $\langle \gamma^H \rangle$ and the similar estimates $\langle
\gamma^V \rangle$ obtained from the {\tt WFPC2} study (CS98; in Figure
2 b the comparison between the average nuclear cusp slopes obtained
from the {\tt F160W} images and those obtained from the {\tt F110W}
images shows excellent agreement between the two estimates, so that
results similar to those discussed for the $H$ band hold when the
$J$-band data are used instead).  There is a small trend for the
galaxies with the steepest NIR cusp slopes to have steeper optical
slopes. Although in principle this could be due to a small nuclear
color gradient in the direction of a bluer center in these systems,
the most likely explanation for the trend is that it comes from a
spurious residual effect of the wider PSF in the NIR images.  The
errors bars are larger than the scatter between points. The major
uncertainty in estimating the nuclear cusp slopes arises from the
unknown shape of the underlying nuclear stellar profile; independently
deriving the cusp slopes in the different wavelength regions leads to
a correlation in the errors if the shapes are intrinsically similar in
the different passbands.  Independent from these considerations, by
far the most striking result that comes out from this plot is the very
good agreement between the average nuclear cusp slopes obtained for
any galaxy in the optical and in the NIR wavelength regions.

Diffuse dust effects cannot in principle be ruled out; however, it is
unlikely that they may conspire so well to reproduce for all systems
the same nuclear cusp slopes in the optical and in the NIR wavelength
region. A more plausible interpretation is that once the patches of
dust and the knots of star formation are properly masked out from the
optical images, both the visual and the NIR passbands provide a fair
representation of the underlying nuclear stellar populations in spiral
galaxies.  The similarity between the visual and NIR average nuclear
cusp slopes has two immediate consequences: First, it implies that
there are small or no global color gradients within the innermost
$\sim 50$-100 pc from the galaxy centers.  In the absence of `tuned'
conspiracies between dust and stellar populations ages/metallicities,
this in turn means that within that region there are no major global
variations in the stellar population properties of nearby spirals, in
contrast with the widespread local color variations documented in
paper I, which imply significant local variations in stellar
populations, star formation rates, and/or dust.  Second, the
similarity implies that the large range in nuclear stellar cusps
observed in both wavelength regions is due to real differences in the
rate of increase of the stellar density rather than to dust or recent
star formation effects.

In Figure 3 we show the average slope $\langle \gamma^H \rangle$ as a
function of the Hubble type of the host galaxy.  There is a clear
trend showing that the nuclear cusp slope decreases with increasing
Hubble type, i.e., late-type galaxies have shallower nuclear cusp
slopes than earlier-type spirals.  A straightforward way to assess the
statistical significance of this trend is to derive the distributions
of $\langle \gamma \rangle$ values separate for each morphological
type, and verify by means of a Kolmogorov-Smirnov test whether these
distributions could be drawn from the same one. The number of objects
per morphological type in our sample is however too small to draw a
firm conclusion. We therefore binned galaxies in only two groups, one
with Hubble types earlier or equal to T=3 and the other with types
later than this `threshold', respectively; this division provides bins
with roughly equal numbers (23 and 21 objects).  The two derived
distributions show a probability of only $\approx 5\%$ of being drawn
from the same parent distribution. A similar probability is obtained
by changing the threshold type of $\pm 1$, although the two bins in
these cases are differently populated (29 and 15 objects for T=2, and
7 and 37 objects for T=4).  The dispersion in the Hubble type versus
$\langle \gamma^H \rangle$ relation is however large, and in
particular galaxies with intermediate types $\sim 3 \pm$1 cover the
entire range of nuclear stellar cusp slope values, from very shallow
to very steep.  The {\tt NICMOS} light profiles are not extended
enough in radius to perform a bulge-disk decomposition; however, the
latter is available for several objects from our previous {\tt WFPC2}
survey.  Using this information on the bulge light profile, in Figure
3 we identify galaxies with different bulge properties: empty squares
represent galaxies with $R^{1/4}$-law bulge light profiles, while
filled squares represent galaxies with exponential bulges. Galaxies
for which no fit to the bulge component is available are represented
by 4-points stars.  The $R^{1/4}$ bulges are confined to the earlier
types and have on average steep cusps $\langle \gamma^H \rangle$, in
contrast with the exponential bulges, which are confined to the
intermediate types and have on average rather shallow stellar
cusps. Interestingly, the only galaxy with a very steep nuclear cusp
among the group with Hubble type $\ge$Sbc is NGC2344, which is also
the only galaxy in that group which hosts an $R^{1/4}$-type bulge.
While it is possible that NGC 2344 has been incorrectly classified, it
is a fact that $R^{1/4}$ and exponential bulges are both found
embedded in the intermediate-type spirals (Courteau et al.\ 1996;
Carollo et al.\ 1998).

The difference in nuclear behaviour between exponential and $R^{1/4}$
bulges is further illustrated in Figure 4a, where we show the $\langle
\gamma^H \rangle$ versus the absolute $V$ bulge magnitude for our
sample of spirals. Symbols for the exponential and $R^{1/4}$ bulges
are as in Figure 3.  The dichotomy between the two classes of bulges
is quite evident at any given luminosity in the range of overlap, with
the exponential bulges showing significantly shallower cusp slopes
than the $R^{1/4}$ bulges.  Within the caveats of the small number
statistics, there is no evidence for galaxies belonging to the
different `nuclear morphological classes' introduced in paper I
(galaxies with concentrated nuclear star formation mixed with dust,
galaxies with diffuse blue nuclear regions, galaxies with regular
nuclear/circum-nuclear dust and galaxies with irregular
nuclear/circum-nuclear dust) occupying any particular region of the
$M_V$ versus $\langle \gamma \rangle$ diagram (Figure 4 b).
Significantly shallower cusp slopes for exponential relative to
$R^{1/4}$ bulges were also found in the {\tt WFPC2} data (CS98);
indeed, this result is a corollary of stating that there is good
agreement between the $\langle \gamma \rangle$ derived in the optical
and in the NIR (Figure 2 a).  The NIR data are, however, important to
make sure that the dichotomy between steep cusp slopes in
$R^{1/4}$-law systems and shallow cusps in exponential bulges is again
not a spurious result driven by the almost ubiquitous presence of dust
and recent star formation in the centers of spirals, but a real
physical effect that extends to the pc-scales the structural
difference that exists between these systems on the kpc-scales.  As
discussed in CS98, this dichotomy is not the result of a trivial
inward extrapolation of the two different analytical forms, since the
$\langle \gamma^{H} \rangle$ values are obtained from a different
radial range than that used to perform the bulge $R^{1/4}$--law or
exponential fits, that was taken $> 1''$.

Shown in Figure 4a as dots are also the measurements obtained for a
sample of elliptical galaxies from a similar {\tt NICMOS} study
conducted by Quillen et al.\ (2000). The comparison between the
ellipticals and the massive $R^{1/4}$-law bulges of spirals shows that
the latter have average nuclear cusp slopes similar to those of
elliptical galaxies of similar luminosity.  Again this was suggested
by the optical data, but it is now unambiguosly demonstrated by the
NIR observations.  The fact that $R^{1/4}$ disk-embedded bulges have
nuclear stellar cusps (and thus densities) indistinguishable from
those of diskless elliptical galaxies of similar luminosities suggests
that for these systems the presence of a large-scale disk, which in
principle may alter the nuclear structure of the embedded spheroid,
has in fact little or no influence on the spheroid.

For the $R^{1/4}$ spheroids, the correlation between the total
spheroid luminosity, as described by $M_V$, and the strength of the
nuclear stellar cusp, as described by $\langle \gamma \rangle$, is not
the only known relationship between a {\it global} and a {\it nuclear}
quantity. The total spheroid luminosity has been found to correlate
also with the mass of the central black hole $M_{BH}$ (e.g., Magorrian
et al.\ 1998); in fact, an even better correlation is found between
the central black hole mass $M_{BH}$ and the {\it mass} of the
spheroid $M_{sph}$, as described by its velocity dispersion $\sigma$;
Ferrarese \& Merritt 2000; Gebhardt et al.\ 2000). These relationships
imply a `triangular' correlation for $R^{1/4}$ spheroids between
$M_{sph}$, $M_{BH}$ and $\langle \gamma \rangle$, drawn schematically
in the bottom panel of Figure 5, which is likely to hold valuable
information on the process of $R^{1/4}$-type spheroid formation.  Our
result that at any given total luminosity (and thus likely mass) of
overlap, the exponential bulges have systematically lower nuclear
stellar cusps than their $R^{1/4}$ relatives may be indicative of a
breakdown of the above `triangular' relationship, as illustrated
schematically in the upper panel of Figure 5.  Indeed, the fact that the
exponential bulges do not lie on the $M_{sph}$-$\langle \gamma
\rangle$ correlation that holds for the $R^{1/4}$ spheroids 
could imply that for these systems at least one of the other two
correlations, either the $M_{BH}$-$\langle \gamma \rangle$, or the
$M_{sph}$-$M_{BH}$, or even both must also break down.  An alternative
is that what we are observing for the exponential bulges is not a
breakdown of the $M_{sph}$-$\langle \gamma
\rangle$ correlation and a consequent breaking of the 
triangular relation among $M_{sph}$-$\langle \gamma \rangle$-$M_{BH}$,
but rather a similar triangular relationship, `displaced' however to a
different region of parameter space.  A `displaced' triangular
relationship for the exponential bulges could explain the $M_{BH}$
estimate for the Milky Way bulge: this is the only exponential bulge
for which the black hole mass has been accurately measured and found
to be significantly smaller than what is expected on the basis of the
$\sigma$-$M_{BH}$ correlation valid for the $R^{1/4}$ spheroids.
Kinematic information for a significant sample of exponential bulges
is still missing, and is awaited to clarify whether or not the
processes that form the exponential bulges produce a global to nuclear
connection of the kind found in the $R^{1/4}$ spheroids.  Either way,
it is clear that either the disappearance or the modification of the
nuclear versus global relationships for the exponential bulges
requires abundant consideration in the context of understanding the
formation of this class of spheroids and their connection if any with
the $R^{1/4}$ spheroids.

\section{Summary}

\noindent
Using the {\tt NICMOS} dataset introduced in Carollo et al.\ (2001,
paper I), we have presented the results of the isophotal fits
performed on the F160W images (56 galaxies) and on the F110W images
(23 objects).  We have modelled the derived surface brightness
profiles with the analytical law introduced by Lauer et al.\ (1995),
and used the smooth analytical representation of the data to compute
the strength of the average nuclear cusp slopes $\langle \gamma
\rangle$.  Our findings about the NIR structural properties of spiral
galaxies are very similar to those that we had obtained by
investigating for the same systems in the optical wavelength
region. This lack of a major difference between the visual and the NIR
analysis of the nuclear stellar density cusps carries important
consequences for our understanding of the nuclear structure of spiral
galaxies, since it establishes that the diversity unveiled in their
central regions is not an artifact of `polluting' effects but the
manifestation of real nuclear differences in systems with different
global properties.

\bigskip

\acknowledgements We thank Phil James for useful discussions,
Roeland van der Marel for providing the pedestal-removal software, and
the anonymous referee for many comments that have improved the
presentation of this work.  This research has been partially funded by
Grants GO-06359.01-95A and GO-07331.02-96A awarded by STScI, and has
made use of the NASA/IPAC Extragalactic Database (NED) which is
operated by the Jet Propulsion Laboratory, Caltech, under contract
with NASA.

\newpage

\begin{table*}
{\tiny\begin{center}\begin{tabular}{lcccccccc}
\hline\hline
\multicolumn{1}{l}{Name} &
\multicolumn{1}{c}{Radial range} &
\multicolumn{1}{c}{$R_b$} &
\multicolumn{1}{c}{$\alpha$} &
\multicolumn{1}{c}{$\beta$} &
\multicolumn{1}{c}{$\gamma$} &
\multicolumn{1}{c}{$\mu_b$} & 
\multicolumn{1}{c}{$\langle \gamma_H \rangle$}  &
\multicolumn{1}{c}{$\langle \gamma_V \rangle$}\\
 &($"$)&($"$)&&&&(mag)&&\\
\hline
N289   &    0.00-5.00    &   0.60    &  12.22    &   0.69   &    0.91   &  15.68   &    -0.82 & -\\
N406   &    0.30-8.00    &   6.12    &   2.81    &   1.46   &    0.12   &  19.22   &    -0.10 & -0.07\\
N488   &    0.00-9.48    &   1.81    &   1.15    &   1.42   &    0.50   &  15.73   &    -0.51 & -0.55\\
N772   &    0.00-8.00    &   0.40    &  11.54    &   1.02   &    0.96   &  14.51   &    -0.89 & - \\
N972   &    0.30-6.67    &   0.55    &  11.25    &   0.82   &    0.40   &  15.06   &    -0.35 & - \\
N1345  &    0.30-8.00    &   4.03    &   2.50    &   3.81   &    0.14   &  19.46   &    -0.12 & -0.10\\
N1398  &    0.00-8.00    &   0.92    &   3.32    &   1.11   &    0.60   &  14.37   &    -0.52 & - \\
N1483  &    0.30-8.00    &   1.78    &   6.12    &   0.32   &    0.12   &  18.90   &    -0.10 & -0.03\\
N1688  &    0.00-8.00    &   2.20    &  12.88    &   0.94   &    0.08   &  17.98   &    -0.06 & - \\
N1892  &    0.00-5.00    &   0.73    &   3.55    &   0.15   &    0.42   &  18.38   &    -0.34 & - \\
N2082  &    0.30-8.00    &   3.10    &   5.21    &   0.65   &    0.37   &  18.62   &    -0.32 & -0.34 \\
N2196  &    0.30-8.00    &   1.78    &   1.14    &   1.36   &    0.65   &  16.14   &    -0.61 & -0.66 \\
N2344  &    0.01-9.81    &   0.12    &   2.76    &   1.04   &    0.84   &  14.35   &    -0.89 & -0.92 \\
N2460  &    0.00-8.71    &   1.27    &   6.83    &   0.95   &    0.81   &  15.82   &    -0.71 & -0.86 \\
N2748  &    0.00-8.00    &   0.86    &   5.01    &   0.71   &    0.46   &  16.31   &    -0.39 & - \\
N2758  &    0.30-8.00    &   1.98    &   3.36    &   1.28   &    0.32   &  18.57   &    -0.27 & -0.08\\
N2903  &    0.00-5.00    &   0.43    &  20.22    &   0.43   &    0.95   &  14.99   &    -0.83 & - \\
N2964  &    0.00-5.00    &   0.78    & 170.37    &   1.32   &    0.88   &  15.37   &    -0.79 & - \\
N3031  &    0.30-8.00    &   1.90    &   0.65    &   1.49   &    0.24   &  13.60   &    -0.43 & - \\
N3067  &    0.00-5.00    &   1.09    &   2.53    &   0.69   &    0.63   &  16.81   &    -0.53 & - \\
N3259  &    0.30-8.00    &   0.00    &   7.33    &   0.68   &    0.44   &  15.60   &    -0.60 & -0.73 \\
N3277  &    0.30-8.00    &   0.68    &  24.54    &   1.28   &    0.92   &  14.99   &    -0.84 & -0.85 \\
N3898  &    0.00-7.77    &   0.99    &   1.00    &   1.67   &    0.50   &  14.50   &    -0.62 & -0.65 \\
N3900  &    0.00-8.82    &   0.15    &   3.15    &   1.15   &    0.87   &  13.61   &    -0.96 & -1.07 \\
N3949  &    0.30-8.00    &   7.44    &   1.53    &   2.57   &    0.26   &  18.39   &    -0.23 & - \\
N4219  &    0.30-8.00    &   0.92    &   4.03    &   1.14   &    0.41   &  15.97   &    -0.36 & -\\
N4384  &    0.00-5.00    &   1.03    &   8.53    &   0.90   &    0.23   &  17.64   &    -0.19 & -0.12 \\
N4527  &    0.00-8.00    &   1.51    &   1.61    &   1.79   &    0.27   &  14.76   &    -0.31 & - \\
N4536  &    0.00-8.00    &   2.43    &   2.13    &   2.06   &    0.91   &  15.76   &    -0.83 & - \\
N5188  &    0.00-5.00    &   0.85    &  17.16    &   1.51   &    0.44   &  15.22   &    -0.37 & - \\
N5488  &    0.00-8.00    &   0.81    &   3.23    &   1.42   &    0.57   &  15.50   &    -0.50 & - \\
N5678  &    0.00-5.00    &   0.82    &   2.02    &   1.02   &    0.94   &  15.87   &    -0.87 & - \\
N5985  &    0.00-7.84    &   1.42    &  19.04    &   0.94   &    0.76   &  17.06   &    -0.67 & -0.72 \\
N6340  &    0.01-9.81    &   0.37    &   3.10    &   1.23   &    0.67   &  14.19   &    -0.70 & -0.79 \\
N6384  &    0.30-8.00    &   2.00    &   0.76    &   1.78   &    0.01   &  16.32   &    -0.26 & -0.25 \\
N7013  &    0.00-8.00    &   1.98    &  13.55    &   1.50   &    0.78   &  15.34   &    -0.69 & - \\
N7162  &    0.30-8.00    &   2.73    &   2.86    &   0.73   &    0.43   &  18.65   &    -0.36 & - \\
N7217  &    0.00-8.00    &   0.46    &   7.00    &   0.92   &    0.39   &  14.21   &    -0.38 & - \\
N7280  &    0.00-8.00    &   0.77    & 137.00    &   1.46   &    1.00   &  14.99   &    -0.93 & -0.93 \\
N7421  &    0.30-8.00    &   2.03    &   4.92    &   0.96   &    0.76   &  17.89   &    -0.66 & -0.75 \\
N7513  &    0.30-8.00    &   2.27    &   1.54    &   0.86   &    0.30   &  17.76   &    -0.27 & - \\
I4390 &    0.00-8.00    &   0.37    &   3.21    &   1.50   &    0.56   &  14.86   &    -0.68  & - \\
I5271 &    0.00-8.00    &   0.36    &   2.70    &   0.81   &    1.14   &  15.17   &    -1.02  & - \\
I5273 &    0.00-5.00    &   0.56    &   8.53    &   0.89   &    0.18   &  16.64   &    -0.18  & - \\
\hline
\end{tabular}\end{center}}
\caption{Results of the Nuker fits to the $H$ surface brightness
profiles.  The columns in the table list, from left to right: the
galaxy name, the radial range of the reported best fit parameters, and
the corresponding best fit values for $R_b$, $\alpha$, $\beta$,
$\gamma$ and $\mu_b=-2.5 \log I_b$. The last column on the right lists
the values of the average nuclear stellar cusp $\langle \gamma^H
\rangle$ computed within 0.1$''$-$0.5''$. When available and for easy reference, 
we report in the last column the $V$-band measurements of the nuclear
stellar cusp slope $\langle
\gamma^V \rangle$ published in CS98.}
\label{tab1}
\end{table*}

\normalsize

\newpage

\begin{table*}
{\tiny\begin{center}\begin{tabular}{lccccccc}
\hline\hline
\multicolumn{1}{l}{Name} &
\multicolumn{1}{c}{Radial range} &
\multicolumn{1}{c}{$R_b$} &
\multicolumn{1}{c}{$\alpha$} &
\multicolumn{1}{c}{$\beta$} &
\multicolumn{1}{c}{$\gamma$} &
\multicolumn{1}{c}{$\mu_b$} & 
\multicolumn{1}{c}{$\langle \gamma_J \rangle$} \\
 &($"$)&($"$)&&&&(mag)&\\
\hline
N406    &       0.30-8.00     &    3.17     &   15.49     &    0.56    &     0.16    &  19.27     &    -0.15  \\
N772    &       0.00-8.00     &    0.44     &    5.42     &    1.02    &     0.97    &  15.13     &    -0.98  \\
N972    &       0.03-5.00     &    1.01     &   95.99     &    0.91    &     0.56    &  16.15     &    -0.51  \\
N1398   &       0.00-8.00     &    0.84     &    2.45     &    1.14    &     0.56    &  14.74     &    -0.54  \\
N2082   &       0.40-5.00     &    2.09     &    1.50     &    0.77    &     0.27    &  18.93     &    -0.27  \\
N2196   &       0.30-8.00     &    2.80     &    0.98     &    1.65    &     0.64    &  17.11     &    -0.68  \\
N3067   &       0.00-5.00     &    1.15     &    3.99     &    0.65    &     0.60    &  17.53     &    -0.55  \\
N3277   &       0.30-8.00     &    0.66     &   32.94     &    1.30    &     0.90    &  15.40     &    -0.88  \\
N3949   &       0.30-8.00     &    5.96     &    1.73     &    2.28    &     0.27    &  18.48     &    -0.25  \\
N4527   &       0.00-8.00     &    1.01     &    2.21     &    1.45    &     0.29    &  15.06     &    -0.33  \\
N5678   &       0.30-8.00     &    0.00     &    8.47     &    0.94    &     0.52    &  14.00     &    -0.93  \\
N5985   &       0.00-7.70     &    1.64     &   25.12     &    0.99    &     0.77    &  17.65     &    -0.73  \\
N7162   &       0.40-5.00     &    2.03     &    1.18     &    0.77    &     0.33    &  18.89     &    -0.34  \\
N7217   &       0.00-8.00     &    0.50     &   10.00     &    0.94    &     0.40    &  14.72     &    -0.39  \\
\hline
\end{tabular}\end{center}}
\caption{Same as in Table 1, but for the $J$-band.}
\label{tab2}
\end{table*}
\normalsize

\newpage

\begin{figure}
\caption{Panel a: The $H$ surface brightness profiles for the 56
galaxies for which an isophotal fit could be performed on the {\tt
NICMOS F160W} images.  The PSF-convolved Nuker best fits, when they
could be obtained, are shown as solid lines superimposed on the data.
Panel b: As in Panel a, but for the 23 galaxies with $J$-band
isophotal fits.}
\end{figure}

\begin{figure}
\caption{Panel a: $V$-band average nuclear slope 
$\langle \gamma^V \rangle$ inside $0.1''$-$0.5''$ reported in CS98
versus the similar measurement in the $H$-band. There is a very good
agreement between the optical and the NIR measurements.  Panel b: As
in Panel a, but the comparison is now between the $H$ and the $J$-band
measurements. There is very good agreement between the two estimates.
}
\end{figure}

\begin{figure}
\caption{Average nuclear slope $\langle \gamma^H_{\rm fit} \rangle$
versus Hubble type (from RC3).  Empty squares are the $R^{1/4}$-law
bulges, solid squares are the exponential bulges (analytical fits to
the bulges were derived in C98). Galaxies with no bulge component, or
for which the shape of the bulge profile could not be modelled from
the optical data are represented by 4-points stars.}
\end{figure}

\begin{figure}
\caption{Panel a: Average nuclear slope $\langle \gamma^H_{\rm fit} \rangle$
versus absolute $V$ magnitude of the spheroidal component. Symbols for
the exponential and $R^{1/4}$ bulges are as in Figure 3; galaxies with
no bulge component, or for which the shape of the bulge profile is
unknown, are excluded from this figure.  The dots represent the
elliptical galaxies from the {\tt NICMOS} study of Quillen et al.\
2000. Panel b: Same as in Panel a, but only for our
measurements. Symbols in this panel identify the four morphological
classes introduced in paper I (filled triangles: galaxies with
concentrated nuclear star formation mixed with dust; open squares:
galaxies with diffuse blue nuclear regions; open pentagons: galaxies
with regular nuclear/circum-nuclear dust; stars: galaxies with
irregular nuclear/circum-nuclear dust).}
\end{figure}

\begin{figure}
\caption{A visualization of the spheroid luminosity -- nuclear stellar
cusp slope $\langle \gamma \rangle$ -- mass of central black hole
`triangular' relationship for the $R^{1/4}$ bulges (lower panel). The
lower $\langle \gamma \rangle$ values for a given luminosity measured
for the exponential bulges imply a `breaking down' of this triangular
relationship (upper panel; see text). }
\end{figure}

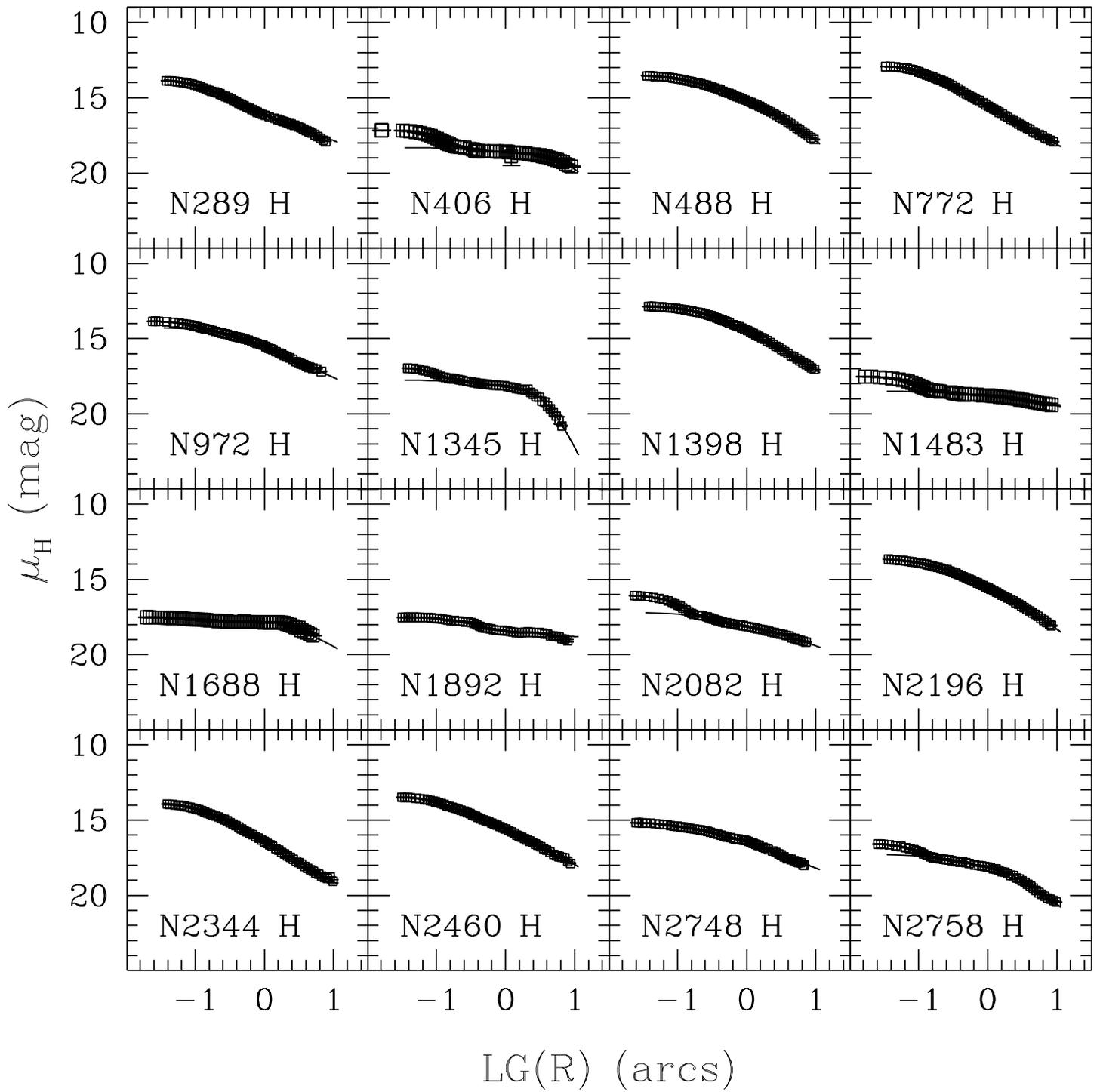
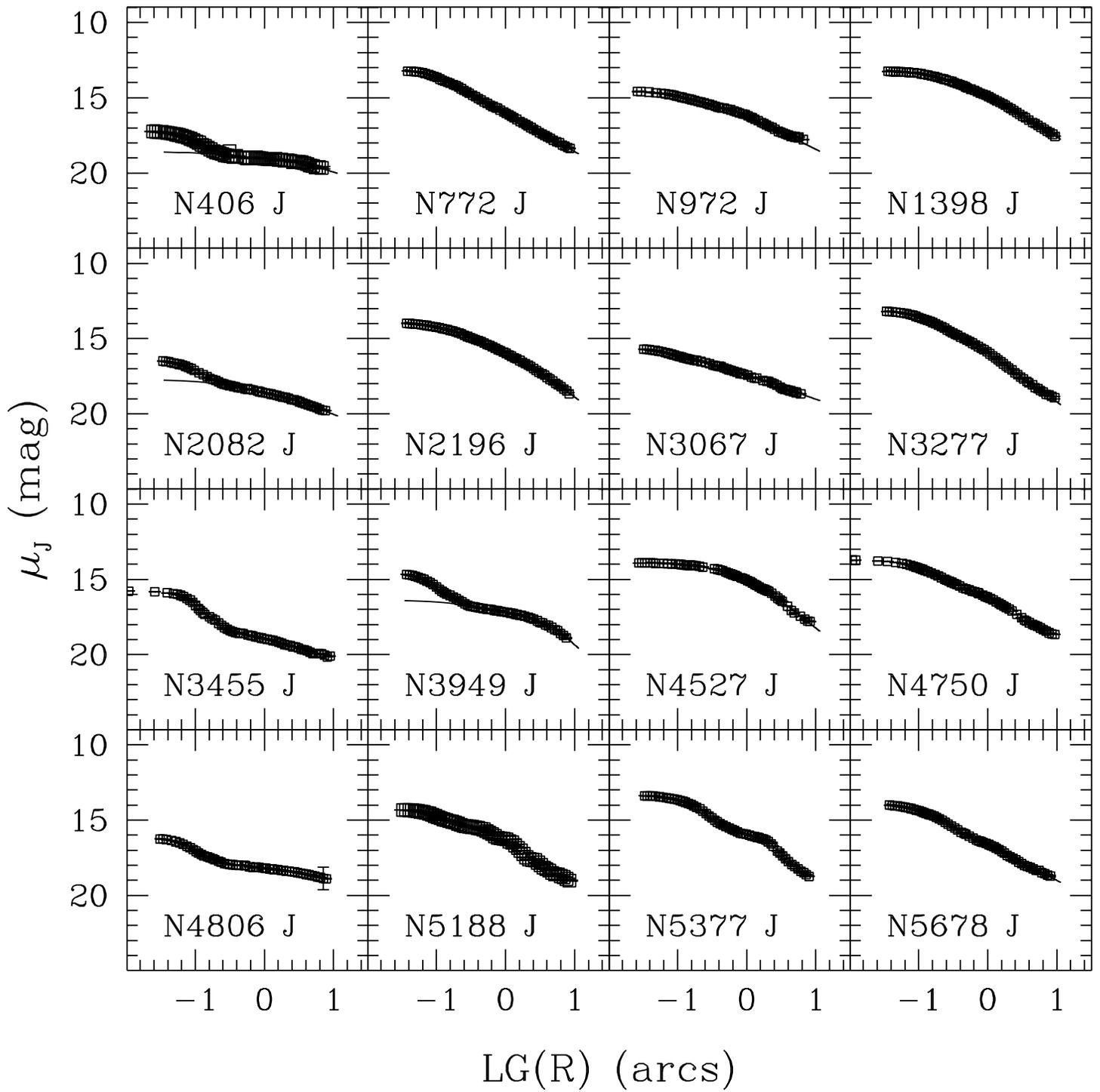
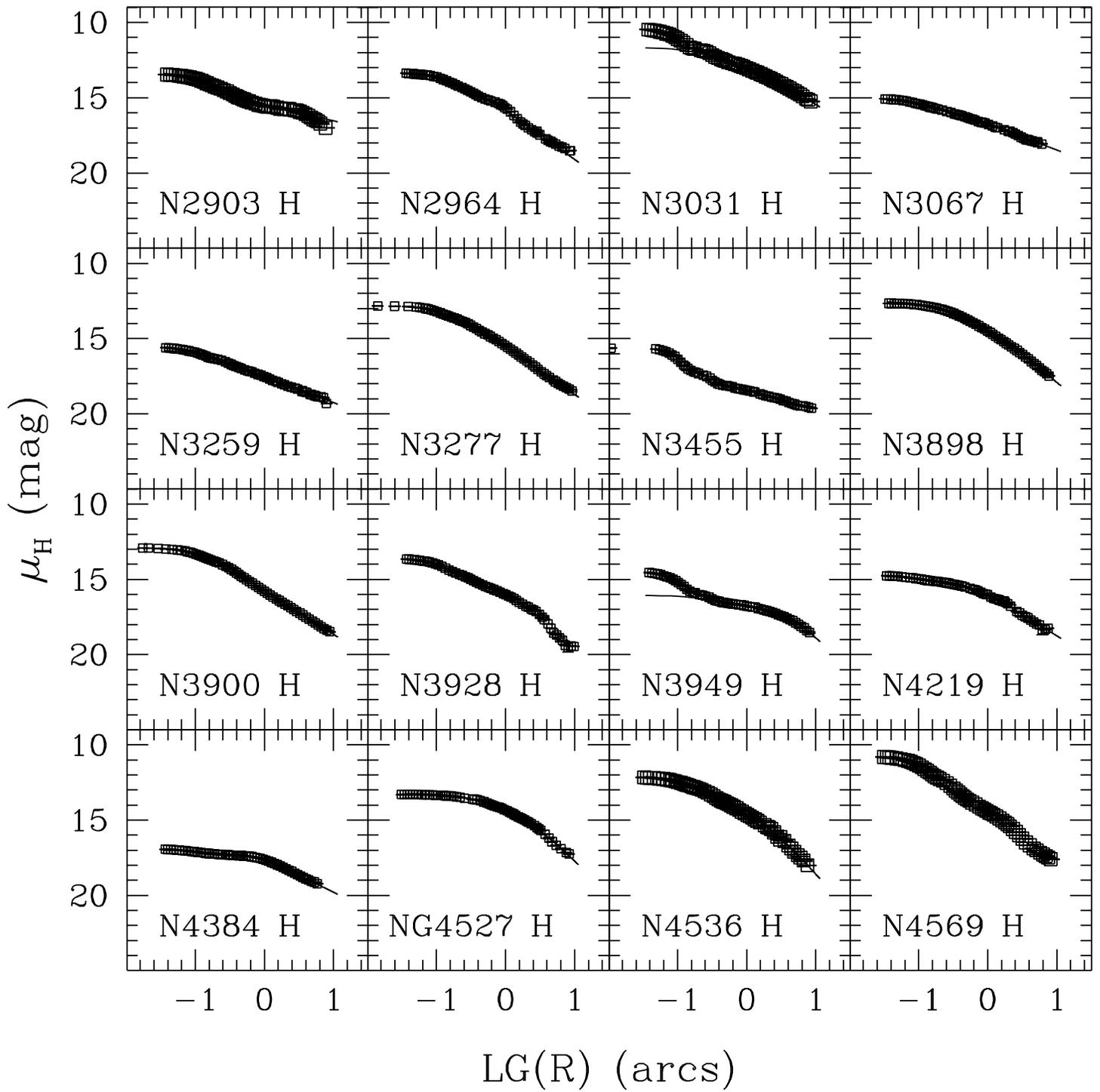
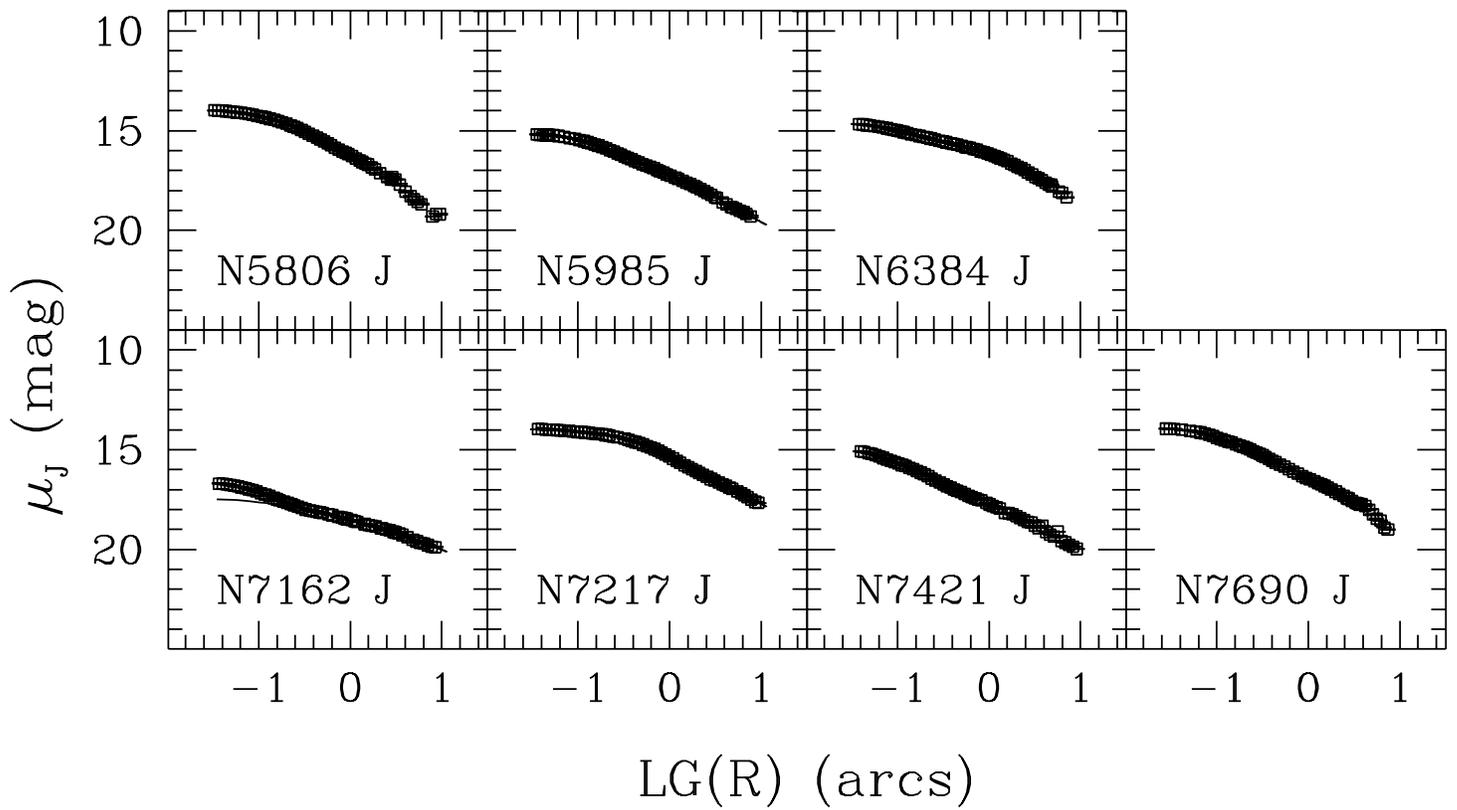
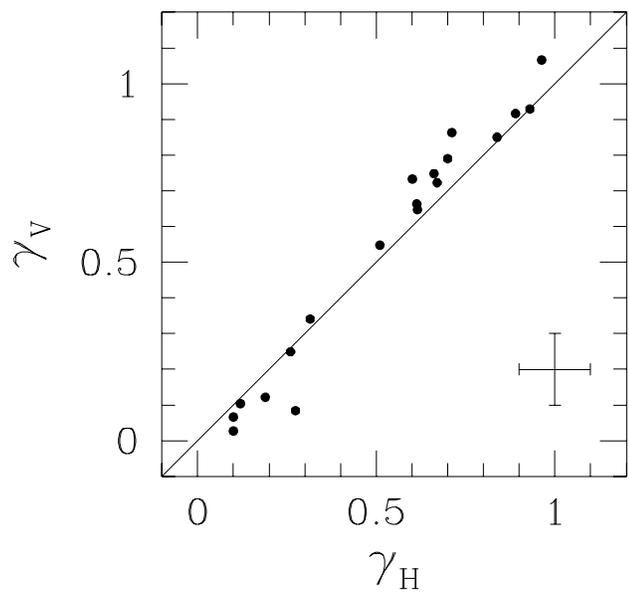
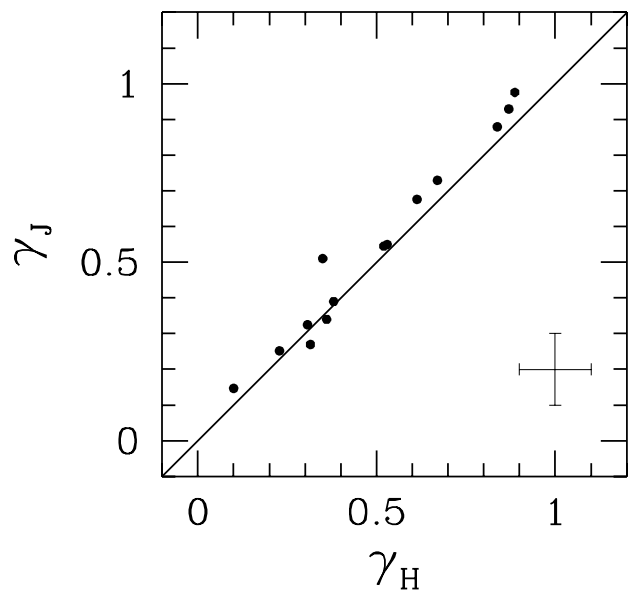
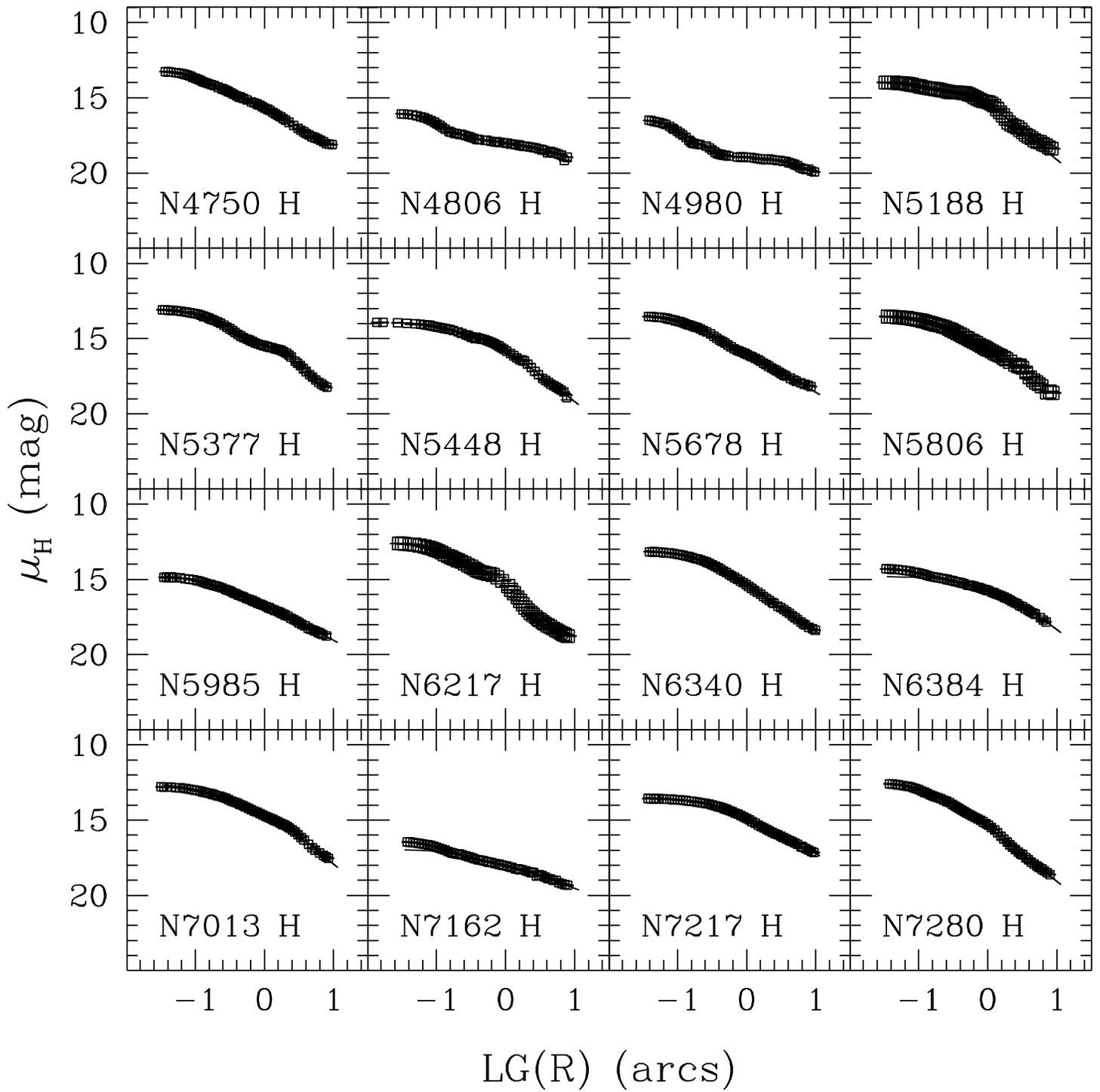
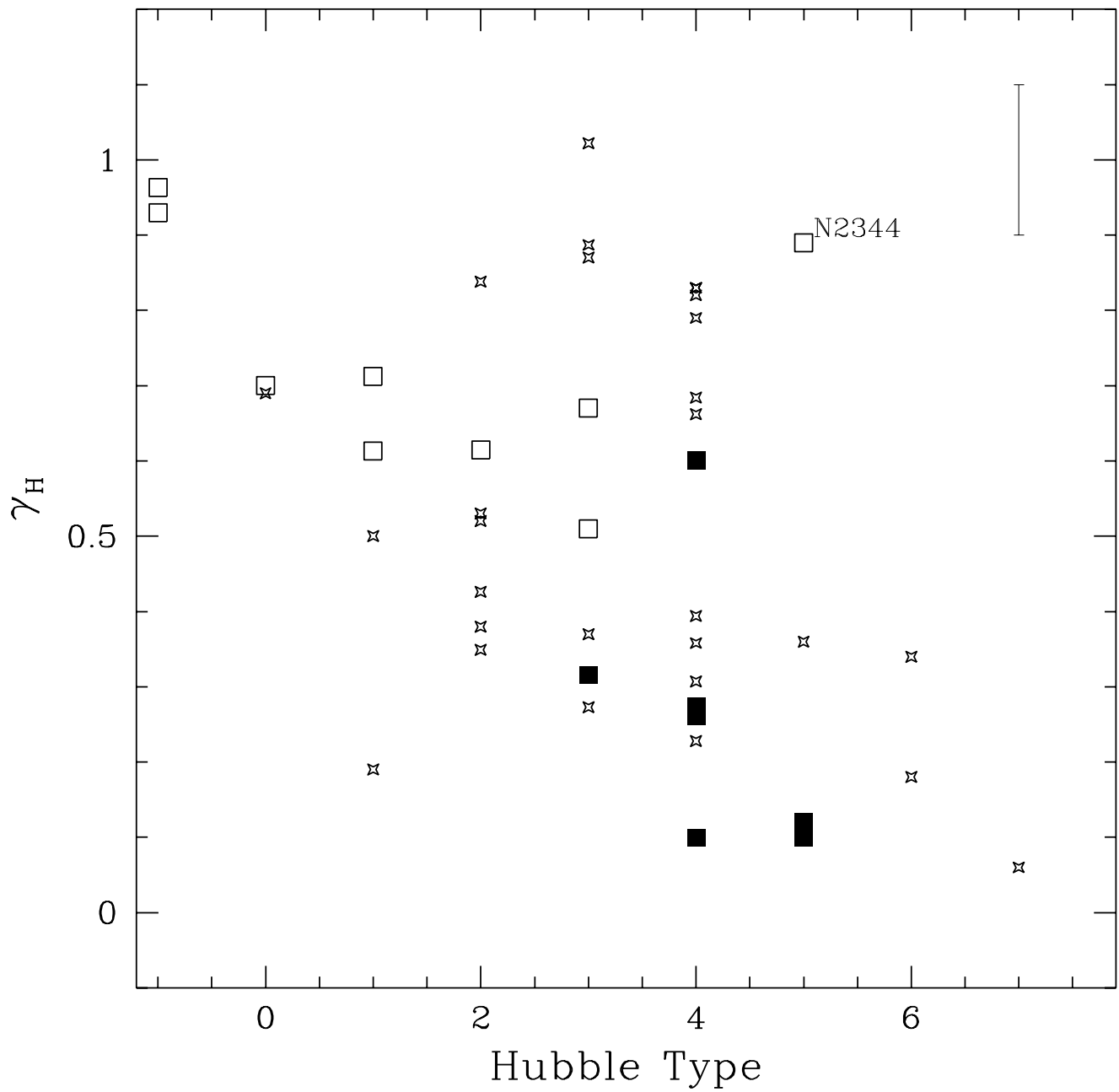
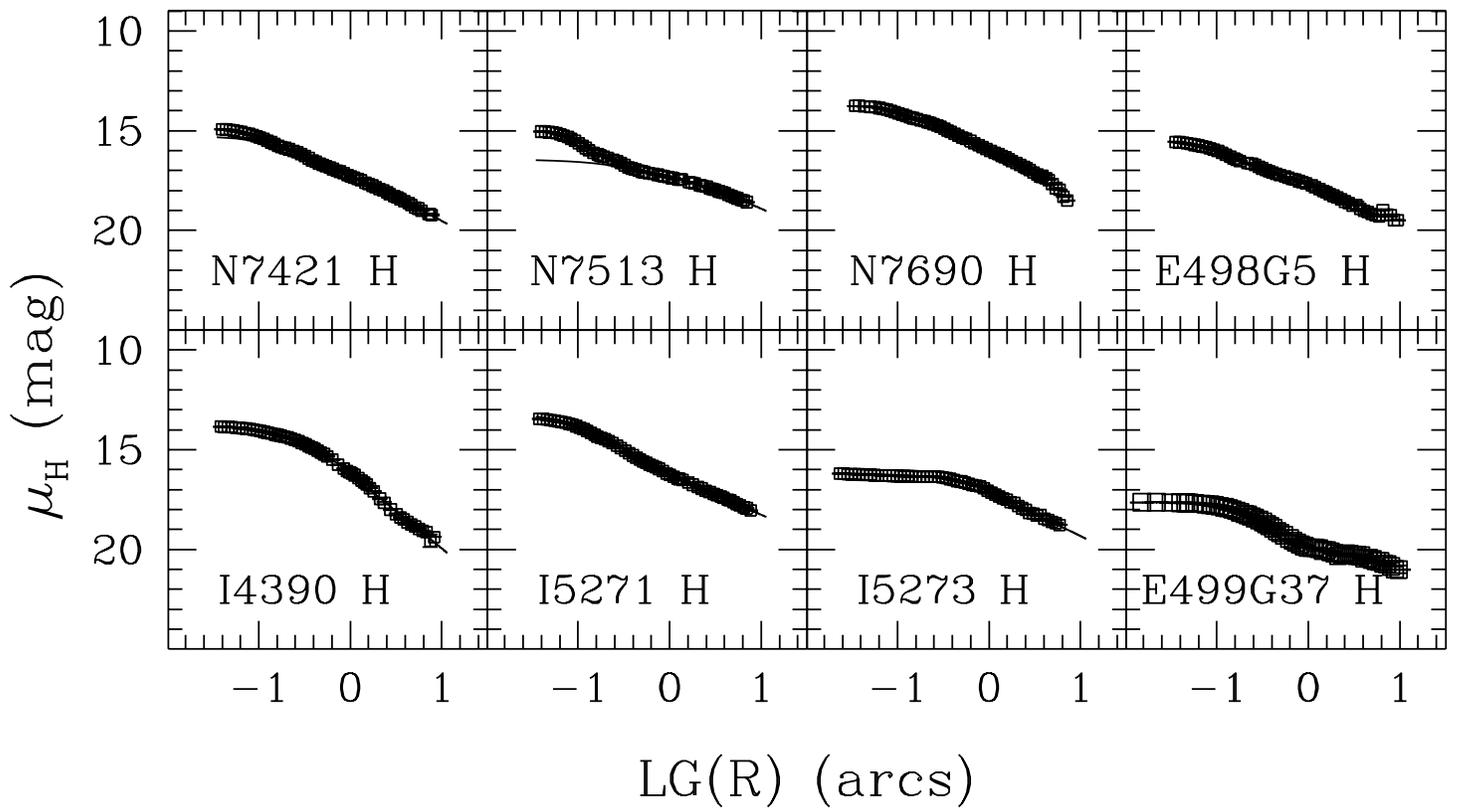
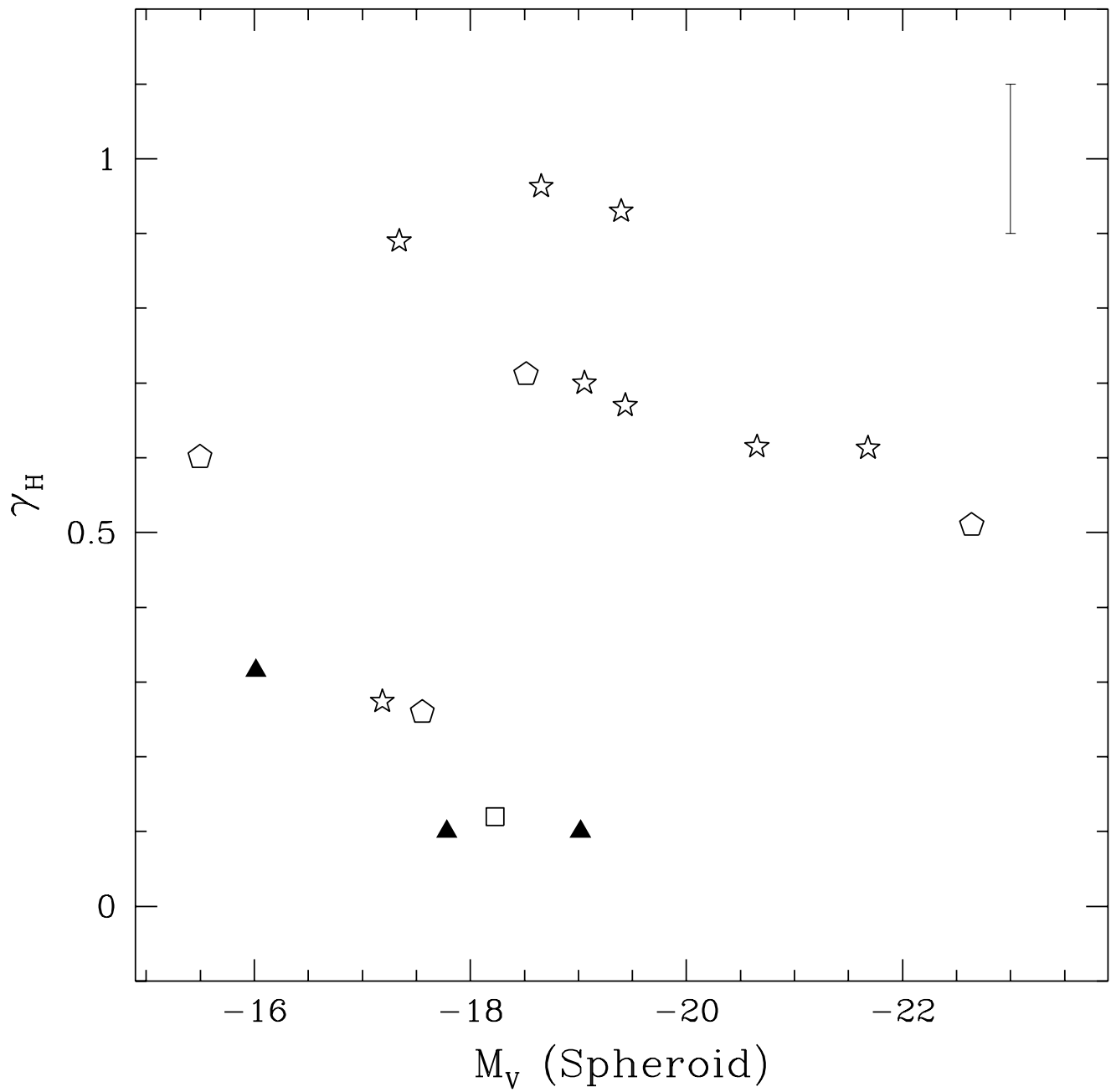
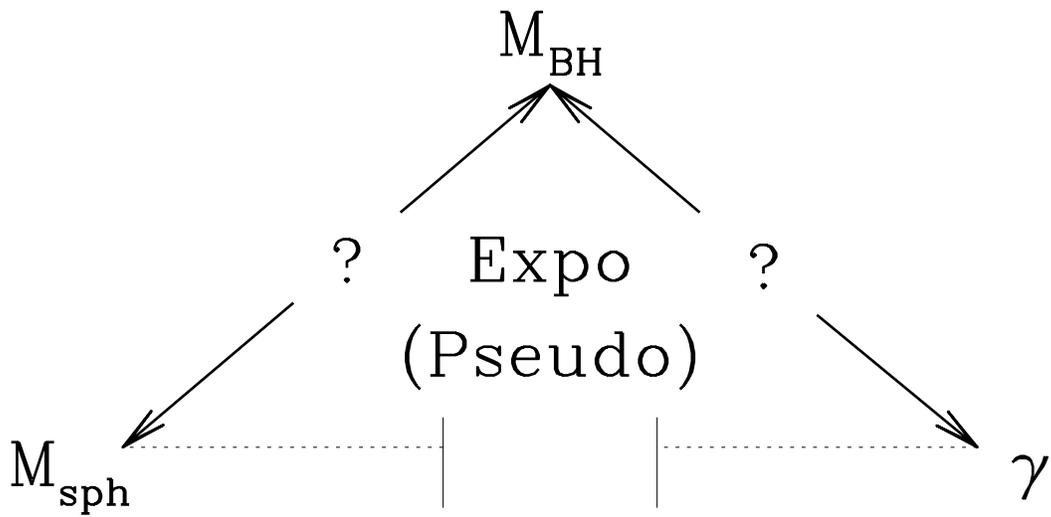
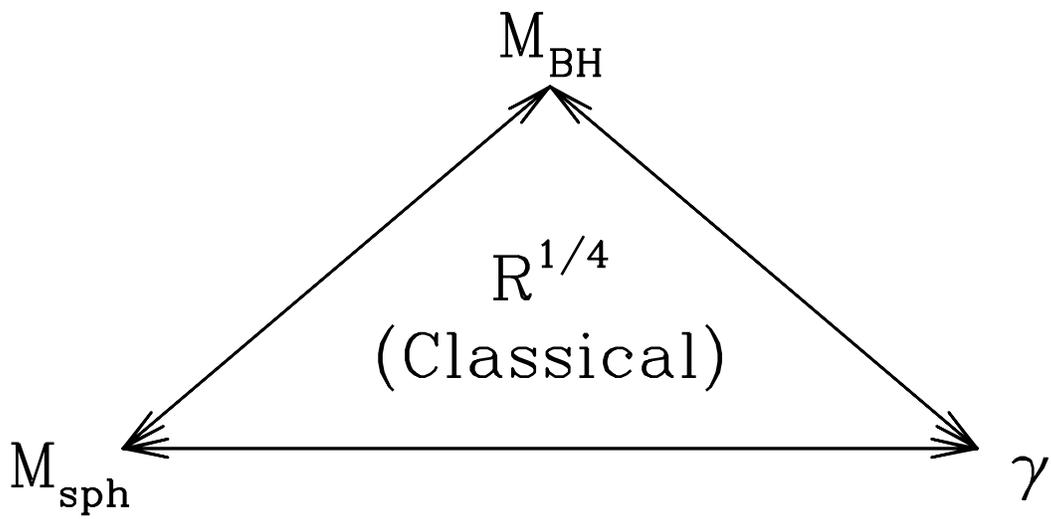
\end{document}